\documentclass{emulateapj}

\usepackage{subfigure}

\usepackage{supertabular}
\usepackage{rotating}
\usepackage{natbib}
\usepackage{graphicx}
\usepackage[space]{grffile}
\usepackage{latexsym}
\usepackage{amsfonts,amsmath,amssymb, amsthm}
\usepackage{url}
\usepackage[utf8]{inputenc}
\usepackage{fancyref}
\usepackage{hyperref}
\hypersetup{colorlinks=false,pdfborder={0 0 0},}
\usepackage{textcomp}
\usepackage{longtable}
\usepackage{multirow,booktabs}
\def\ha{H$\alpha~$}
\def\hap{H$\alpha$}
\def\iha{$I_{\textrm{H}\alpha}~$}
\def\ihap{$I_{\textrm{H}\alpha}$}
\def\lha{$[L_{\textrm{H}\alpha}/L_{\textrm{bol}}]~$}

\def\id{$I_{\textrm{D}}~$}
\def\idp{$I_{\textrm{D}}$}

\def\ica{$I_{\textrm{\ion{Ca}{1}}}~$}

\def\io{I$_2~$}

\def\kep{\emph{Kepler}~}

\def\md{M\,dwarf~}
\def\mdp{M\,dwarf}
\def\mds{M\,dwarfs~}
\def\mdsp{M\,dwarfs}
\def\na{\ion{Na}{1}~}
\def\nap{\ion{Na}{1}}
\def\caone{\ion{Ca}{1}~}
\def\ca{\ion{Ca}{2}~}
\def\mps{m\,s$^{-1}$~}
\def\mpsp{m\,s$^{-1}$}
\def\actaa{Acta~Astronomica~}

\shorttitle{Stellar Activity and its Implications for Exoplanet Detection on GJ 176}
\shortauthors{Robertson et al.}

\begin{document}

\title{Stellar Activity and its Implications for Exoplanet Detection on GJ 176}

\author{Paul~Robertson$^{1,2,3}$, Michael~Endl$^{3}$, Gregory~W.~Henry$^{4}$, William~D.~Cochran$^{3}$, Phillip~J.~MacQueen$^{3}$, and Michael~H.~Williamson$^{4}$
\\
\normalsize{$^{1}$Department of Astronomy and Astrophysics, The Pennsylvania State University, University Park, PA 16802, USA; pmr19@psu.edu}\\
\normalsize{$^{2}$Center for Exoplanets \& Habitable Worlds, The Pennsylvania State University}\\
\normalsize{$^{3}$Department of Astronomy and McDonald Observatory, University of Texas at Austin, Austin, TX 78712, USA}\\
\normalsize{$^{4}$Center of Excellence in Information Systems, Tennessee State University, 3500 John A. Merritt Blvd., Box 9501, Nashville, TN 37209, USA}
}

\begin{abstract}
We present an in-depth analysis of stellar activity and its effects on radial velocity (RV) for the M2 dwarf GJ 176 based on spectra taken over 10 years from the High Resolution Spectrograph on the Hobby-Eberly Telescope.  These data are supplemented with spectra from previous observations with the HIRES and HARPS spectrographs, and $V$- and $R$-band photometry taken over 6 years at the Dyer and Fairborn observatories.  Previous studies of GJ 176 revealed a super-Earth exoplanet in an 8.8-day orbit.  However, the velocities of this star are also known to be contaminated by activity, particularly at the 39-day stellar rotation period.  We have examined the magnetic activity of GJ 176 using the sodium I D lines, which have been shown to be a sensitive activity tracer in cool stars.  In addition to rotational modulation, we see evidence of a long-term trend in our \na D index, which may be part of a long-period activity cycle.  The sodium index is well correlated with our RVs, and we show that this activity trend drives a corresponding slope in RV.  Interestingly, the rotation signal remains in phase in photometry, but not in the spectral activity indicators.  We interpret this phenomenon as the result of one or more large spot complexes or active regions which dominate the photometric variability, while the spectral indices are driven by the overall magnetic activity across the stellar surface.  In light of these results, we discuss the potential for correcting activity signals in the RVs of \mdsp.
\end{abstract}

\section{\bf Introduction}

\mds are currently highly desirable targets for exoplanet surveys, as they allow for detection of terrestrial, potentially habitable planets with current or upcoming technology.  In addition to our own \md survey \citep{endl03,endl06}, virtually every exoplanet search program now dedicates a significant portion of its time allotment surveying M stars \citep[e.g.][]{haghighipour10,bonfils13,berta13}.  An outstanding problem surrounding the discovery of the lowest-mass planets is that below RV amplitudes of $\sim 5$ \mpsp, stellar activity will cause velocity shifts that may imitate or distort the signal of an exoplanet \citep{queloz01,huelamo08,dumusque12,robertson13b,santos14}.  This problem is especially acute for \mdsp, since the magnetic activity of old M stars has not been studied as thoroughly as for old solar-type stars.

Nearby M stars are of immense interest for exoplanet discovery and characterization.  Statistics of \kep planets suggest terrestrial-size planets should be common around M stars \citep{dressing13}, and the observational advantages of M star planets (relatively high RV amplitudes and planet-to-star radius ratios) mean M dwarfs in the solar neighborhood will offer the earliest opportunities to characterize potentially Earthlike worlds.  As a result, upcoming RV instruments such as CARMENES \citep{quirrenbach14}, HPF \citep{mahadevan14}, and SPIRou \citep{artigau14} will focus intensely on these nearby cool stars.  

To date, the most compelling exoplanets found with RV around M stars orbit very magnetically quiet M dwarfs such as GJ 581 and GJ 667C \citep[although even those stars exhibit significant RV contributions from magnetic activity; see][]{robertson14a,robertson14b}.  On the other hand, new dedicated M dwarf RV surveys will target nearby mid-late M stars, which tend to be more rapidly rotating and magnetically active.  In a photometric activity survey of \kep targets, \citet{basri13} find that the fraction of \emph{all} M stars more active than the Sun is much higher than for hotter stars, exceeding 90\% in some temperature bins.  Some studies have suggested that the RV amplitudes of stellar signals may be reduced in the infrared \citep{reiners10,marchwinski14}.  However, even low-amplitude noise from activity will be problematic for identifying habitable zone super-Earths.  More effort must therefore be invested in understanding activity in late-type stars and its effects on RV.

\begin{table*}
\begin{center}
\begin{tabular}{| l l l |}
\hline & & \\
Stellar Parameter & Value & Reference \\
\hline & & \\
Spectral Type & M2 & \citet{vonbraun13} \\
$V$ & $9.951 \pm 0.012$ & \citet{koen10} \\
$K$ & $5.607 \pm 0.034$ & \citet{koen10} \\
Parallax & $107.83 \pm 2.85$ mas & \citet{vanl07} \\
Proper Motion & $\mu_{\alpha} = 656.85 \pm 3.81$ mas/yr & \\
 & $\mu_{\delta} = -1116.20 \pm 2.49$ mas/yr & \citet{vanl07} \\
Distance & $9.27 \pm 0.24$ pc & \\
Mass & $0.50 \pm 0.02~M_{\odot}$ & \citet{delfosse00} \\
Radius & $0.4525 \pm 0.0221~R_{\odot}$ & \citet{vonbraun13} \\
$T_{eff}$ & $3679 \pm 77$ K & \citet{vonbraun13} \\
Luminosity & $0.0337 \pm 0.0018~L_{\odot}$ & \citet{vonbraun13} \\
Metallicity ([M/H]) & $0.07 \pm 0.15$ & \citet{schlaufman10} \\
\hline
\end{tabular}
\caption{Stellar parameters for GJ 176}
\label{stellartab}
\end{center}
\end{table*}

GJ 176 is an excellent archetype of a planet-host M star with activity levels in between those of very quiet stars, such as GJ 581, and the more active mid-M dwarfs targeted by upcoming experiments.   \citet{endl08} initially claimed detection of a $24~M_{\oplus}$ planet in a 10.2-day orbit around the star, but \citet{butler09} showed that this orbital solution is inconsistent with the Keck/HIRES RV data.  The HIRES data did show high RV variability for the star, suggesting an exoplanet might still exist in the system.  \citet{forveille09} presented a 2-component solution to their HARPS RV data, including a low-mass planet ($M \sin i = 8.3~M_{\oplus}$) on an 8.8-day orbit and an RV signal near 40 days, which analyses of photometry \citep{kiraga07}, \hap, and the \ca H\&K lines showed is the stellar rotation period.  As a nearby ($d = 9.3$pc) \md with one known super-Earth already, GJ 176 is a very attractive target for follow-up searches for additional low-mass planets.  However, conclusively demonstrating the presence of any further planets in this system will require a thorough understanding and treatment of the stellar magnetic activity.

In this article, we examine the magnetic activity of GJ 176 using spectral and photometric activity indicators.  In addition to the previously-observed rotation signal, we see evidence of a very long-term activity trend which drives a slope in the observed radial velocities.  We explore to what degree this information may be used to increase the detection efficiency for planets in the system.

\section{\bf Stellar Properties}

At a distance of 9.3 parsecs, GJ 176 (M2) is among a handful of known exoplanet hosts within 10 parsecs.  With a $V$-band magnitude of $\sim10$, it is of roughly average brightness among our \md RV targets.  We list the complete set of stellar properties for GJ 176 in Table \ref{stellartab}.  For the mass and metallicity, although more recent estimates exist for this star, we use the photometric calibration techniques of \citet{delfosse00} and \citet{schlaufman10} to remain consistent with the stellar characterization presented in our full-sample activity survey of \mds \citep{robertson13a}.

\section{\bf Data}

\subsection{Radial Velocity}

We have carried out a dedicated \md radial velocity survey using the High Resolution Spectrograph \citep[HRS;][]{tull98} on the Hobby-Eberly Telescope \citep[HET;][]{ramsey98}.  Full details on the survey and its results, including the observing strategy and target properties can be found in \citet{endl03,endl06} and \citet{robertson13a}.  The HRS acquires high-precision RVs via the iodine cell technique, in which an \io absorption cell is placed in the light path to superimpose thousands of weak, stable \io absorption lines over the target spectrum.  These lines serve as a wavelength reference, allowing us to model the stellar Doppler shift necessary to produce the observed star-plus-\io spectrum.  Our RVs are extracted using the AUSTRAL software package \citep{endl00}.

We have obtained 98 RVs for GJ 176 over 10 years, which we list in Table \ref{rvtab}.  We note that earlier velocities based on the same spectra were originally published in \citet{endl08}, but as we have re-reduced our entire data set with our latest version of AUSTRAL, their values may have changed slightly.

GJ 176 has also been observed intensively by the HARPS and HIRES spectrographs.  Our RV analysis in \S\ref{sec:rv} is supplemented by HARPS RVs from \citet{forveille09} and \citet{gds12} and HIRES RVs from \citet{butler09}.

\subsection{Activity-Sensitive Absorption Lines}

Although RV surveys typically monitor the magnetic activity of their targets with the \ca H\&K lines, the wavelength coverage of HRS in our standard RV mode does not extend to those lines.  Instead, we use the \ha and \na D lines as activity tracers for our \md targets.  \ha has been used extensively to study activity in low-mass stars \citep[e.g.][]{kruse10,bell12}, while the use of the \na D resonance feature has more recently been shown to be useful as a cool-star activity tracer \citep{andretta97,diaz07a,gds11}.  We have recently performed activity analyses of our entire \md sample using the \ha \citep{robertson13a} and \na lines, and found that while both indices are helpful for studying magnetic activity, the sodium lines are much more sensitive to activity that causes RV shifts.  In \citet{robertson13b}, we showed that a magnetic cycle in GJ 328 causes RV shifts that cause the orbit of its giant planet to appear more circular than suggested by the activity-corrected velocities.  For the case of GJ 176, where it is desirable to correct activity-related RV signals to identify low-mass planets, the \na D lines are particularly advantageous.

\begin{table*}
\begin{center}
\footnotesize

\begin{tabular}{| l l l l l l |}
\hline & & & & & \\
BJD - 2450000 & RV (\mpsp) & \id (1 \AA~Window) & \id (0.5 \AA~Window) & \iha & \ica \\
\hline & & & & & \\
2935.80776169 & 	 $ -16.44 \pm 4.66 $ & 	 $ 0.12488 \pm 0.00329 $ & 	 $ 0.10040 \pm 0.00236 $ & 	 $ 0.07098 \pm 0.00107 $ & 	 $ 0.02367 \pm 0.00074 $ \\
2939.79788804 & 	 $ -7.46 \pm 6.05 $ & 	 $ 0.12582 \pm 0.00327 $ & 	 $ 0.10184 \pm 0.00240 $ & 	 $ 0.07041 \pm 0.00106 $ & 	 $ 0.02368 \pm 0.00072 $ \\
2941.98273285 & 	 $ -0.46 \pm 5.38 $ & 	 $ 0.12042 \pm 0.00315 $ & 	 $ 0.09647 \pm 0.00215 $ & 	 $ 0.07232 \pm 0.00115 $ & 	 $ 0.02368 \pm 0.00070 $ \\
3254.93830390 & 	 $ -0.92 \pm 5.67 $ & 	 $ 0.11368 \pm 0.00390 $ & 	 $ 0.08739 \pm 0.00292 $ & 	 $ 0.07385 \pm 0.00107 $ & 	 $ 0.02396 \pm 0.00080 $ \\
3297.80620165 & 	 $ -25.22 \pm 5.62 $ & 	 $ 0.12046 \pm 0.00272 $ & 	 $ 0.10108 \pm 0.00224 $ & 	 $ 0.07223 \pm 0.00119 $ & 	 $ 0.02392 \pm 0.00074 $ \\
\hline
\end{tabular}

\caption{Radial velocities and spectral activity indices for our HET/HRS spectra of GJ 176.  The full table will be provided as an online-only supplement to the article.}
\label{rvtab}	
\end{center}
\end{table*}

We define our sodium index, \idp, according to the definition of \citet{diaz07a}.  Specifically, we take the ratio of the fluxes inside windows centered on each of the \na D lines ($\lambda_{\textrm{D1}} = 5895.92$ \AA, $\lambda_{\textrm{D2}} = 5889.95$ \AA), divided by the flux in the nearby pseudocontinuum.  We measure \id using both 1 \AA~and 0.5 \AA~windows, since \citet{gds11} show \id more frequently correlates with the $S_{\textrm{HK}}$ \ca index when measured with a 0.5 \AA~window.  In general, we see very little difference in the results from \id between the window sizes, and use the 1 \AA~window unless specifically noted otherwise.  However, there are some minor differences which we will discuss in later sections.  For reference, we include both values of \id alongside their corresponding RVs in Table \ref{rvtab}.

While we are primarily interested in the sodium feature, we have also computed \ihap, the \ha activity index, for each of our HRS spectra.  The procedure we use to measure \iha and its uncertainty is documented fully in \citet{robertson13a}; like \idp, \iha is simply the ratio of the flux in a 1.6 \AA~window centered on the \ha line to the nearby pseudocontinuum.

As a control quantity, we measure the flux index \ica for the \caone line at $\lambda = 6572.795$ \AA.  This line is not sensitive to stellar activity, and should therefore remain roughly constant.  All of our absorption-line indices may be found alongside the RVs in Table \ref{rvtab}.

To enable comparison, we have extracted \id and \iha values (where applicable) from the HARPS and HIRES spectra discussed in \citet{forveille09} and \citet{butler09} via the ESO\footnote{Based on data obtained from the ESO Science Archive Facility under request number 103236.} and Keck public archives, respectively.  We also consider activity indices from the HARPS M dwarf activity survey \citep{gds12}.

Unfortunately, the \na D feature lies on the edge of a spectral order on HIRES, so we cannot obtain \id for the HIRES data.  At the edge of the order, the blaze function is at a minimum, so any recovered \id values would be very low S/N, and the pseudocontinuum against which we measure the index would be even more unreliable.  Furthermore, barycentric velocity shifts frequently cause one or both \na lines to fall off the CCD completely, making \id measurements impossible.

\subsection{Photometry}
\label{sec:photometry}

Prompted by the presence of significant stellar activity in its RVs, we have monitored GJ~176 for photometric variability with the Tennessee State University (TSU) automated Celestron C-14 telescope. The telescope was equipped with an SBIG STL-1001E CCD camera observing through Cousins $R$ and Johnson $V$ filters. During the 2007--2008 observing season, the C14  was located at Vanderbilt Univerity's Dyer Observatory in Nashville, Tennessee, where it collected 426 observations of GJ~176 in the Johnson $V$ pass band.  In 2010, the telescope was relocated to TSU's automated telescope observing site at Fairborn Observatory in the Patagonia Mountains of southern Arizona \citep{ehf2003}.  There, it collected 42 Johnson $V$ observations in the 2010--2011 observing season, 137 $V$ and 87 Cousins $R$ observations in 2011--2012, and 153 $R$ observations during 2012--2013.

\begin{sidewaysfigure*}
\begin{center}
\subfigure[\label{phot_5panel}]{\includegraphics[width=0.45\columnwidth,trim=0 0 0 12cm]{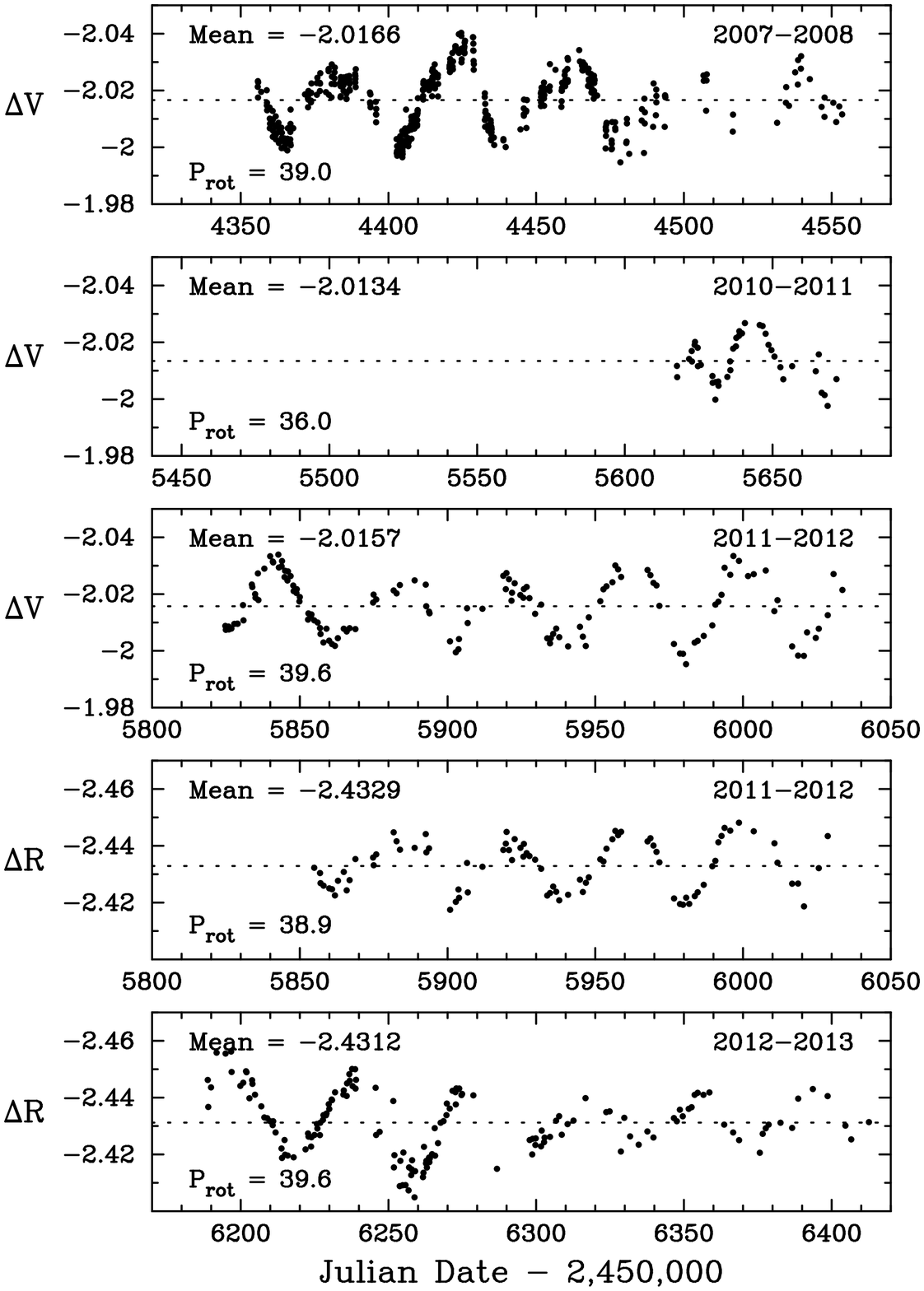}}
\subfigure[\label{phot_comb}]{\includegraphics[width=0.45\columnwidth,trim=0 0 0 12cm]{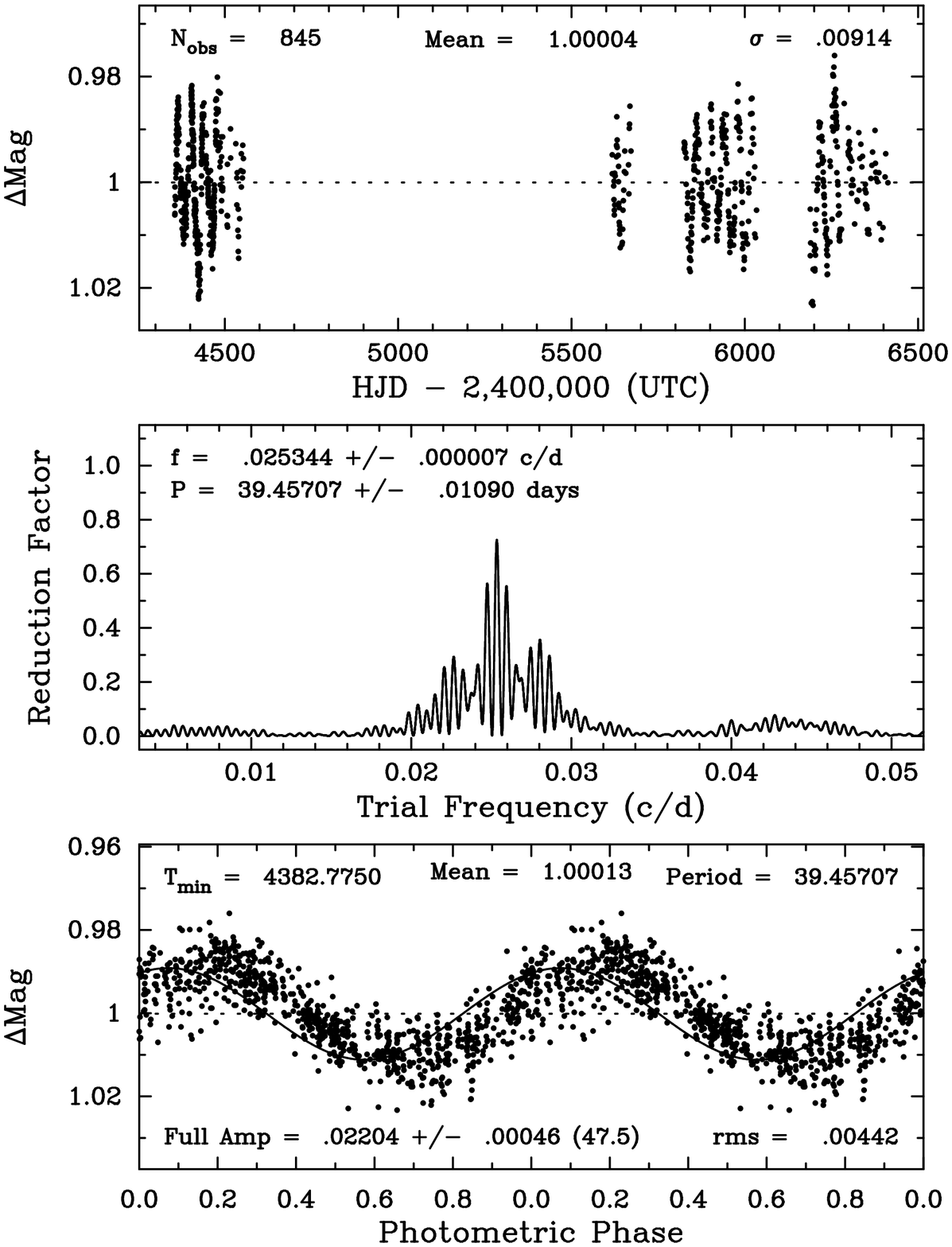}}
\caption{\emph{a}. Differential $V$- and $R$-band photometry of GJ 176.  The top three panels show the data taken in the $V$ band from Dyer (top panel) and Fairborn (panels 2 \& 3) Observatories.  The bottom two panels show the $R$ band data from Fairborn Observatory.  The middle of the each observing season (opposition) occurs roughly around the center of the x axis.  The season mean brightness is indicated by the dotted line in each panel. \emph{b}. Data from \emph{a}, normalized and combined into a single data set (\emph{top}).  The periodogram for the complete set is given in the middle panel.  In the bottom panel, we fold the data to the rotation period, with the best-fit sinusoidal model to the data shown as a solid curve.}
\label{photometry}
\end{center}
\end{sidewaysfigure*}

The $R$-band and $V$-band differential magnitudes are listed in Tables \ref{Rtab} and \ref{Vtab}, respectively, and plotted as a function of Julian date in Figure \ref{photometry}.  Each observation has been corrected for bias, flat-field, differential extinction, and pier-side offset. The differential magnitudes are computed against the mean of 4 constant comparison stars identified in the same CCD field of view. Each differential magnitude is the mean of 4 to 10 successive frames taken on a given night.  Period analyses of the $R$-band and $V$-band observations are shown in Figures \ref{Rplot} and \ref{Vplot}.

\begin{table}
\begin{center}
\footnotesize

\begin{tabular}{| l l |}
\hline & \\
BJD - 2450000 & d$R$ \\
\hline & \\
5854.8932 & 	 $ -2.43229 \pm .00044 $ \\
5856.8425 & 	 $ -2.43041 \pm .00061 $ \\
5857.0246 & 	 $ -2.42685 \pm .00087 $ \\
5858.0320 & 	 $ -2.42596 \pm .00026 $ \\
5859.9542 & 	 $ -2.42502 \pm .00063 $ \\
\hline
\end{tabular}

\caption{Differential $R$-band photometry of GJ 176.  Magnitudes are reported relative to the mean of five constant stars in the same CCD field (see Section \ref{sec:photometry}).  The full table will be provided as an online-only supplement to the article.}
\label{Rtab}	
\end{center}
\end{table}

\begin{table}
\begin{center}
\footnotesize

\begin{tabular}{| l l |}
\hline & \\
BJD - 2450000 & d$V$ \\
\hline & \\
4355.7962 & 	 $ -2.02130 \pm .00110 $ \\
4355.8337 & 	 $ -2.02330 \pm .00090 $ \\
4355.8823 & 	 $ -2.01750 \pm .00100 $ \\
4355.9249 & 	 $ -2.02290 \pm .00140 $ \\
4356.9144 & 	 $ -2.01950 \pm .00350 $ \\
\hline
\end{tabular}

\caption{Differential $V$-band photometry of GJ 176.  Magnitudes are reported relative to the mean of five constant stars in the same CCD field (see Section \ref{sec:photometry}).  The full table will be provided as an online-only supplement to the article.}
\label{Vtab}	
\end{center}
\end{table}

\section{\bf Stellar Activity}
\label{sec:activity}

\begin{figure}
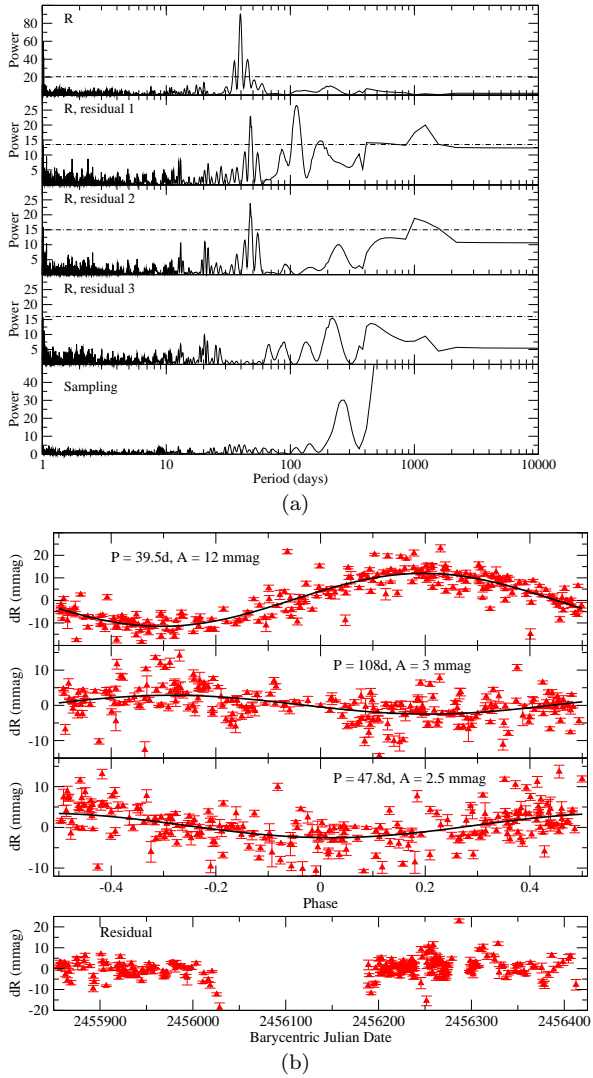

\begin{center}
\subfigure[\label{R_ps}]{\includegraphics[width=0.9\columnwidth,trim=0 0 -1.5cm 0]{gj176R_ps_v2.eps}}
\subfigure[\label{R_curve}]{\includegraphics[width=0.9\columnwidth]{gj176_lightcurvesR_v2.eps}}
\caption{\emph{a}. Generalized Lomb-Scargle periodograms for our $R$-band photometry.  The top panel gives the periodogram of the original data, while the middle three panels give the power spectrum after subtracting a sinusoidal fit to the 40-day stellar rotation period (residual 1), rotation plus the 112-day periodicity (residual 2), and rotation, the 108-day signal, and the 47-day signal (residual 3).  The fifth panel shows the window function, or the periodogram of our time sampling.  The dash-dotted lines indicate the power required for a false alarm probability of 0.01 according to our bootstrap FAP estimate.  \emph{b}. Phase plots of the periodic signals identified in (\emph{a}), and the residuals around the 3-signal fit (\emph{bottom}).  Our sinusoidal fits to the data are given as black lines.}
\label{Rplot}
\end{center}
\end{figure}

We begin our stellar activity analysis by searching for periodic behavior in the photometry and/or the spectral line indices.  In order to examine each data set in its entirety, we use the generalized Lomb-Scargle periodogram as described in \citet{zk09}.  Adapted from the Fourier transform power spectrum, the Lomb-Scargle periodogram excels at identifying periodic behavior in data taken with uneven time sampling.  The generalized version from Zechmeister \& K\"urster additionally allows for individually-weighted data points and floating means.

Because the power of a peak in a Lomb-Scargle periodogram is related to, but not directly indicative of its statistical significance, we estimate false-alarm probabilities (FAPs) for our candidate signal detections using the bootstrap resampling technique of \citet{kurster97}.  The method retains the time stamps of the original data set, while drawing at random (with replacement) a value for each time from the set of observed values.  The periodogram is computed for a large set of such ``fake" data sets, and the FAP is taken as the number of resampled periodograms with a peak at any period stronger than that of the candidate signal.  For each signal discussed below, we have computed a FAP in this manner and listed it in Table \ref{faptab}.

\begin{table}
\begin{center}
\begin{tabular}{| l l l |}
\hline & & \\
Period (days) & FAP & FAP \\
\hline & & \\
\multicolumn{3}{| c |}{Photometric Signals} \\
 & $R$ band & $V$ band \\
\hline & & \\
39.5 & $< 10^{-4}$ & $< 10^{-4}$ \\
112 & $< 10^{-4}$ & $\cdots$ \\
47.4 & $< 10^{-4}$ & $\cdots$ \\
18 & $\cdots$ & $< 10^{-4}$ \\
70 & $\cdots$ & $< 10^{-4}$ \\
\hline & & \\
\multicolumn{3}{| c |}{\id Signals} \\
 & 1 \AA~window & 0.5 \AA~window \\
\hline & & \\
73 & $0.0019$ & $< 10^{-4}$ \\
112 & $0.0055$ & $\cdots$ \\
\hline
\end{tabular}
\caption{False alarm probabilities (FAPs) for periodic signals observed in our data.  FAPs are based on $10^4$ bootstrap resampling trials, as described in \citet{kurster97}.}
\label{faptab}
\end{center}
\end{table}

\begin{figure}
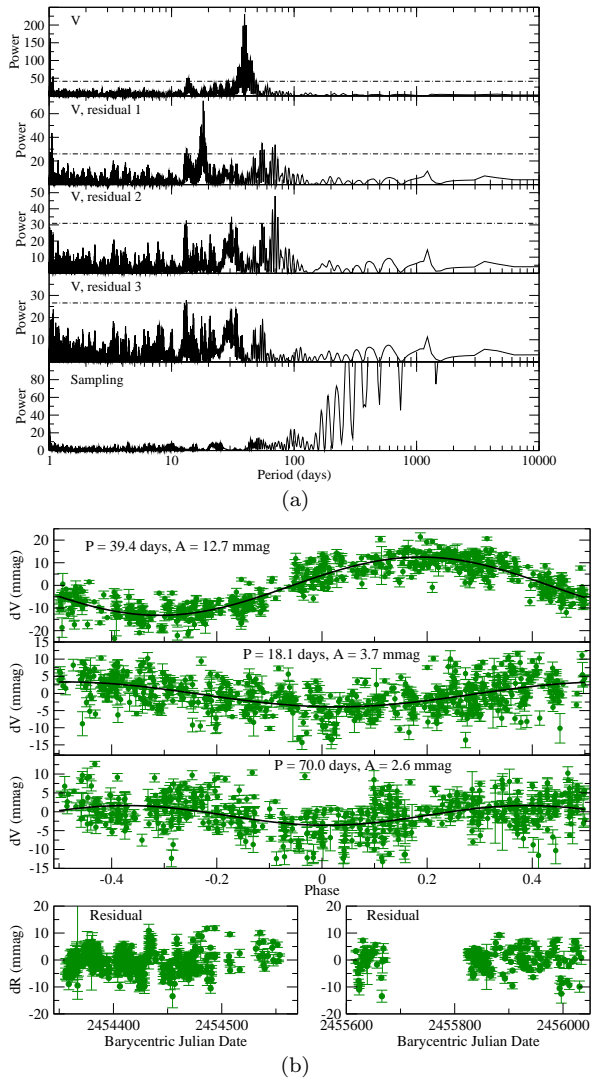

\begin{center}
\subfigure[\label{V_ps}]{\includegraphics[width=0.9\columnwidth,trim=0 0 -1.5cm 0]{gj176V_ps.eps}}
\subfigure[\label{V_curve}]{\includegraphics[width=0.9\columnwidth]{gj176_lightcurvesV_v2.eps}}
\caption{\emph{a}. Generalized Lomb-Scargle periodograms for our $V$-band photometry.  The top panel gives the periodogram of the original data, while the middle three panels give the power spectrum after subtracting a sinusoidal fit to the 40-day stellar rotation period (residual 1), rotation plus its 18-day harmonic (residual 2), and rotation, the 18-day harmonic, and the 70-day signal (residual 3).  The fifth panel shows the window function, or the periodogram of our time sampling.  The dash-dotted lines indicate the power required for a false alarm probability of 0.01 according to our bootstrap FAP estimate.  \emph{b}. Phase plots of the periodic signals identified in (\emph{a}), and the residuals around the 3-signal fit (\emph{bottom}).  Our sinusoidal fits to the data are given as black lines.}
\label{Vplot}
\end{center}
\end{figure}

We show the results of our $R$- and $V$-band photometry in Figures \ref{Rplot} and \ref{Vplot}, respectively.  In both bandpasses, we see a strong periodogram peak near 39.5 days, which has previously been shown to be the rotation period of GJ 176 \citep{kiraga07,forveille09}.  In each bandpass, we have fit a sinusoid of the form $F(t) = F_0 + A\sin(\omega t + \phi)$, where $F_0$ is the mean value, $A$ the amplitude, $\omega = \frac{2\pi}{P}$ the angular frequency, and $\phi$ is the phase.  In $R$, the rotation signal has a period of $39.61 \pm 0.07$ days and an amplitude of $11.7 \pm 0.5$ millimags.  In $V$, the period decreases to $39.44 \pm 0.01$ days, while the amplitude is $12.0 \pm 0.4$ millimags.

In order to obtain the most precise estimate of the rotation period possible, we normalized the $R$- and $V$-band photometry to a mean value of $1.0$ and combined them into a single data set, which we show in Figure~\ref{phot_comb}.  The combined photometry yields a single coherent signal, for which we derive a period of $39.457 \pm 0.011$ days.  A phase plot of the combined photometry at the rotation period is shown in the bottom panel of Figure~\ref{phot_comb}.

Upon fitting and subtracting the 39-day signal, we see a number of additional signals in the residual photometry.  The $R$-band residuals show peaks near 112 days and 47 days, while the $V$-band data show an 18-day harmonic of the rotation period.  After removing the 18-day signal, the $V$-band data show a third periodogram peak at 70 days.  We include fits to these periods in Figures \ref{R_curve} and \ref{V_curve}.  When including the 47-day signal in our model of the $R$-band data, the period of the longest signal shifts slightly, to 108 days.  However, since the periodicity appears consistently at 112 days in multiple indicators, we will continue to refer to it as a 112-day signal.  We note that our residuals to a 3-sinusoid fit to the $V$-band photometry contain some marginally significant power between 10 and 13 days.  Attempting to include a fourth sinusoid in our model does not significantly improve the fit, and there is no clearly preferred period in this range.  We therefore conclude that this excess power is likely noise.

Of the four periods identified in the residual photometry, two can be easily explained.  We interpret the 18-day harmonic as evidence that spots or spot complexes occasionally appear at opposing longitudes on the stellar surface, creating periodicity at half the rotation period.  The 47-day signal is close to both the rotation period and its 1-year alias ($44.2$~days).  It is possible that differential rotation creates periodicities near the fundamental period, as suggested for the false-positive exoplanet signal for HD~41248 \citep{santos14}.  The 70 and 112-day signals, on the other hand, are somewhat surprising.  Rotating starspots are expected to create signals at the rotation period and its integer ratios \citep[i.e.~$P_{rot}/2$, $P_{rot}/3$\ldots,][]{boisse11}, so signals longer than the rotation period but shorter than is typical for magnetic cycles are puzzling.  We will discuss these periodicities further in later sections.

\begin{figure*}
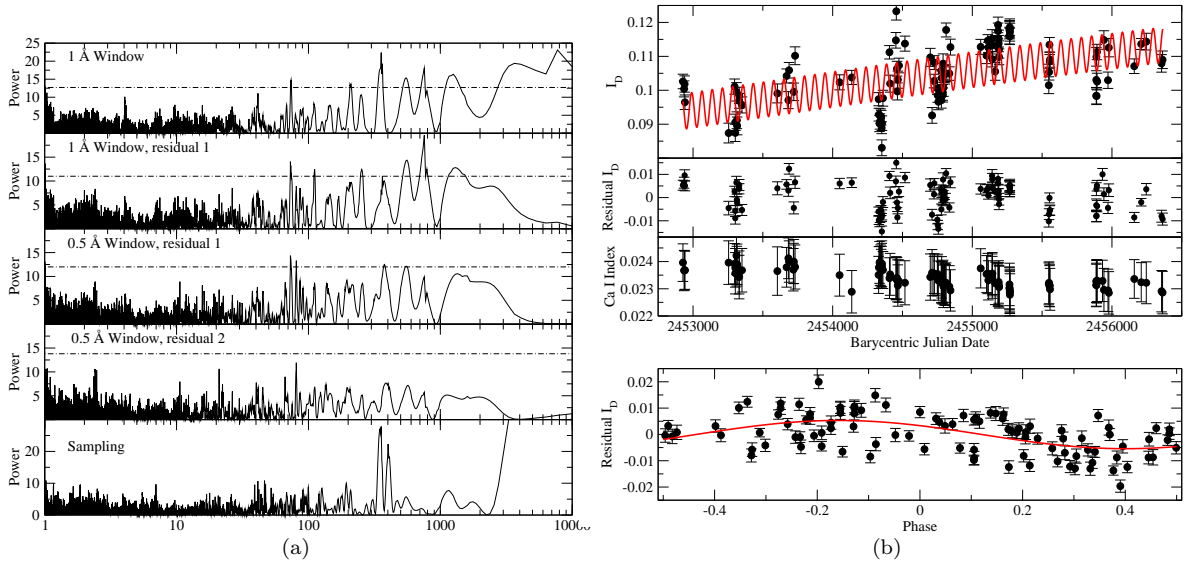

\begin{center}
\subfigure[\label{naps}]{\includegraphics[width=0.9\columnwidth]{gj176_na_pscomp_v2.eps}}
\subfigure[\label{na_fit}]{\includegraphics[width=0.9\columnwidth]{gj176_nai_3.eps}}
\caption{\emph{a}. Generalized Lomb-Scargle periodograms for \idp, our \na D index.  The top panel shows the periodogram of the original time series.  While the power spectrum shown is for the computation of \id using 1 \AA~line windows, we note that the periodogram of the 0.5 \AA~index shows no significant difference.  The two subsequent panels show the residual periodogram after fitting and removing a linear trend from the original data.  Here, we include the periodograms from each of the two \id indices to show that the 73-day period is the only peak that remains constant in each.  The residual periodogram to a trend-plus-sinusoid fit of the 0.5 \AA~\id and the window function are included in the bottom panels.  The horizontal lines indicate the power required for a false alarm probability of 0.5 (dot), 0.1 (dash), and 0.01 (dash-dot) according to our bootstrap FAP estimate.  \emph{b}. \emph{Top}: Time-series \id values from our RV survey.  Our model of the linear trend plus a 73-day sinusoid is given as a solid red line.  Below each \id value is the residual around the 2-signal fit and the corresponding Ca I index, which is an activity-insensitive line used as a control.  \emph{Bottom}: Residual \id values after removing the linear slope, folded to the 73-day period identified in the periodograms.  We note that while the 73-day peak appears in both \id indices, we show the 0.5 \AA~values here because the period appears at higher signal-to-noise.}
\label{na}
\end{center}
\end{figure*}

The periodogram for GJ 176's \id series (Figure \ref{naps}) shows a number of significant peaks.  We discard peaks near 1 and 2 years as aliases caused by our observing cadence; fits to those periods reveal large phase gaps, as is typical of sampling-related aliases.  More compelling is the presence of significant power at longer periods.  While it is possible to fit a complete cycle with $P = 1200$ day sinusoid, the RMS around the fit is essentially equal to that of a straight-line fit, causing us to adopt a linear trend with slope $(5 \pm 1) \times 10^{-6}$~day$^{-1}$ as our best model to the data.

Regardless of the model adopted for the long-term behavior of the \na feature, an additional periodogram peak remains in the residuals at $P = 73$ days.  A bootstrap FAP test produced no false positives in $10^4$ iterations, leading us to conclude the signal is statistically significant.  Furthermore, the period of the peak is an excellent match to the 70-day signal found in our $V$-band photometry.  Since 73 days is close to twice the 39-day rotation period, we speculate this signal may be related to the stellar rotation, although again, such behavior is not predicted by simple spot models.

In analyzing the residual signals in \idp, we find a small dependence on the size of the windows used to compute the index.  While the 73-day peak appears in the periodogram regardless of the window used, the S/N is slightly higher when using 0.5 \AA~windows.  On the other hand, the 1 \AA~windows show some power near the 112-day period observed in the $R$-band photometry, although this detection is less significant.  Because of this slight discrepancy, we show the \id periodograms from both windows in Figure \ref{naps}.  Because of the higher S/N, though, we compute our fit to the 73-day signal using the 0.5 \AA~windows.  When including a sine curve model alongside the linear trend, we find a fit with $P = 73.4 \pm 0.4$d and $A = 0.0052 \pm 0.0001$ (4\%).  

\begin{figure*}
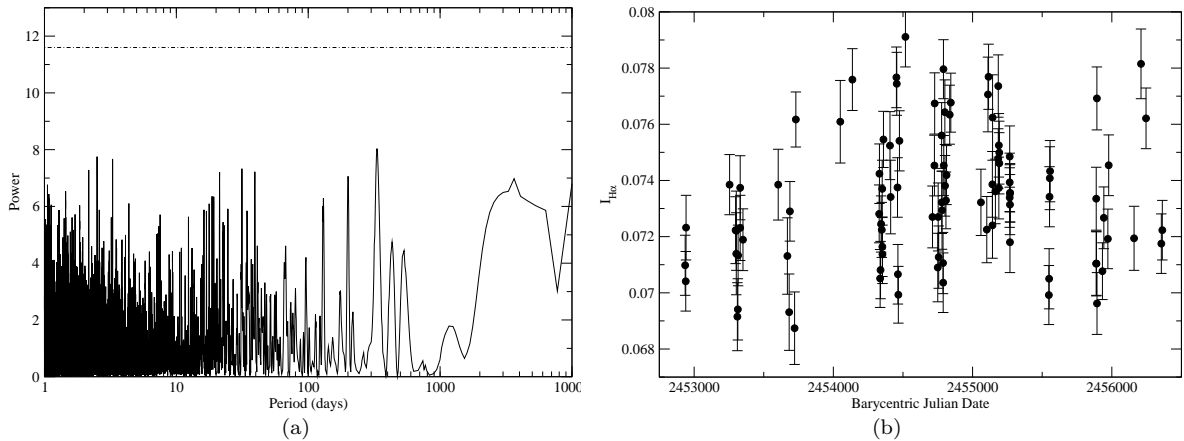

\begin{center}
\subfigure[\label{haps}]{\includegraphics[width=0.9\columnwidth]{gj176_hai_ps.eps}}
\subfigure[\label{ha_time}]{\includegraphics[width=0.9\columnwidth]{gj176_ha.eps}}
\caption{\emph{a}. Generalized Lomb-Scargle periodograms for \ihap, our \ha index.  The dash-dotted line indicates the power required for a false alarm probability of 0.01 according to our bootstrap FAP estimate.  \emph{b}. Our time-series \iha data.  We see no periodic signals over the entire set of our observations in \hap.}
\label{ha}
\end{center}
\end{figure*}

Unlike the photometry and the sodium index, \iha appears to be devoid of coherent periodic behavior.  We show the \iha time series and its periodogram in Figure \ref{ha}.  It is well established \citep[e.g.][]{cincunegui07,gds11} that \ha often does not correlate with other spectral activity tracers, so the absence of the long-term slope or the rotation signal does not imply a fault in our data or analysis.  However, since \citet{forveille09} recovered the 39-day rotation period in \hap, it is important that we point out our overall data set does not show the same signal.

\begin{figure*}
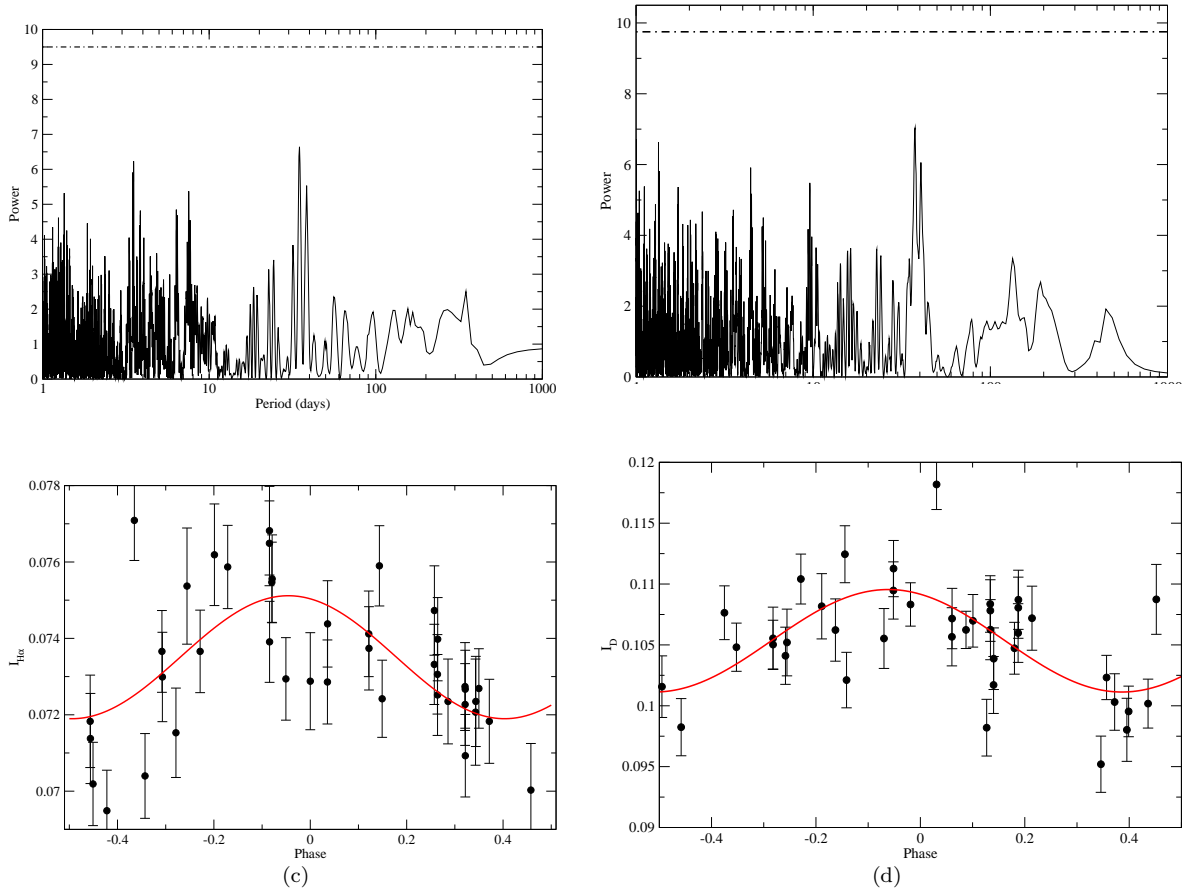

\begin{center}
\subfigure[\label{ha_subset_ps}]{\includegraphics[width=0.9\columnwidth,trim=0 0 -1.5cm 0]{gj176_hai_short_ps.eps}}
\subfigure[\label{na_subset_ps}]{\includegraphics[width=0.9\columnwidth]{gj176_nai_short_ps.eps}}
\subfigure[\label{ha_35d}]{\includegraphics[width=0.9\columnwidth,trim=0 0 -1.5cm 0]{gj176_hai_35d_v2.eps}}
\subfigure[\label{ha_37d}]{\includegraphics[width=0.9\columnwidth]{gj176_nai_37d.eps}}
\caption{We have examined the subset of our spectra spanning September 2008 and March 2010 to search for evidence of the stellar rotation signal in \iha and \idp.  Periodograms of both \iha (\emph{a}) and \id (\emph{b}) show clear peaks near the 39-day rotation period.  The dash-dotted lines indicate the power required for a false alarm probability of 0.01 according to our bootstrap FAP estimate.  We include phase plots of the rotation signal in \emph{c} and \emph{d}.  The presence of the rotation signal in this truncated data (but not in the complete data) suggests the signal does not remain in phase over the duration of our observations.}
\label{subset}
\end{center}
\end{figure*}

It is curious that \id should show a multiple of the rotation period instead of the period itself.  It is also potentially disconcerting that the rotation period does not appear to be present in \iha either, since \citet{forveille09} detected it in both \ca H\&K and \hap.  Examining \iha and \id from the Forveille spectra, we confirm that both \ha and \na D show a periodic signal at the rotation period.  Furthermore, we note that the 112-day periodicity we observe in $R$ appears in the HARPS spectra in \ha and \ca H\&K \citep[][Fig.~6]{forveille09}.

The HARPS data cover a relatively short time span compared to our own, and it is possible that the rotation signal does not stay constant in phase for the absorption-line indices.  We have attempted to verify this hypothesis by restricting the analysis of our data to a subset of dense time sampling from September 2008 to March 2010.  While this time frame is still considerably longer than a single rotation, it is the shortest time over which we have enough data to perform meaningful frequency analysis.  In order to properly evaluate short-period behavior, we have also removed the long-term trend from the \id data.

In Figure \ref{subset}, we show periodograms for \iha and \id over this abbreviated time period.  Here again, we use the 0.5 \AA~window for the \na index because of its slightly higher S/N.  Interestingly, both indices show distinct peaks near the 39-day rotation period, each at a power level similar to the absorption-line detections of \citet{forveille09}.  We fit sinusoids to these peaks, which we include in Figure \ref{subset}, finding best-fit solutions at $P = 34.8$ days (\hap) and $P = 37.5$ days (\nap).  We note, however, that there are a number of periods between 35 and 40 days that are consistent with these data, since the fit is not well constrained with only 40 points.  Based on these detections, we conclude that the rotation period \emph{is} present in our absorption-line indices, but does not remain constant in phase, leading to non-detections when examining the data in its entirety.

\begin{figure*}
\begin{center}
\includegraphics[width=1.55\columnwidth]{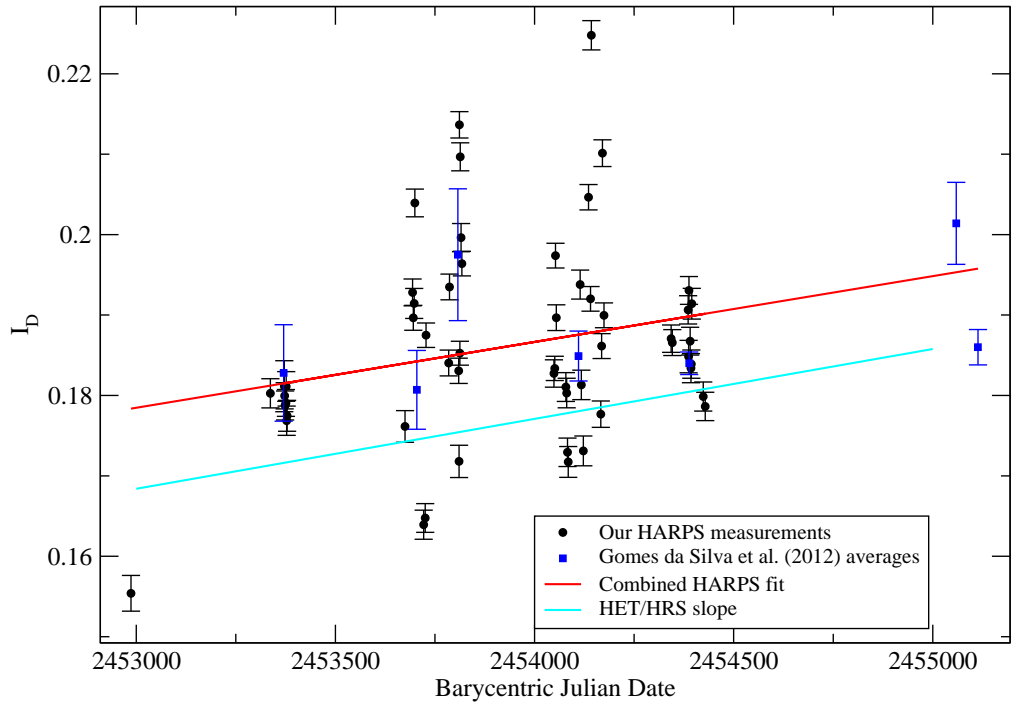}
\caption{\id from the HARPS spectra of \citet[][our measurements, black]{forveille09} and \citet[][blue]{gds12}.  The measured linear trend (red) is consistent with that seen in the HRS data (cyan).  The zero-point offset between the HARPS and HRS slopes is caused by uncertainty in the normalization of \id used by HARPS, and has no physical significance.}
\label{harps_na}
\end{center}
\end{figure*}

While the baseline of the HARPS observations make assessing the presence of the long-term \id trend difficult to assess conclusively, we note that the HARPS \id values are consistent with the slope we observe with HRS.  In Figure \ref{harps_na}, we show our measurements of \id from the \citet{forveille09} spectra alongside those measured by \citet{gds12}.  We have scaled our values by a factor of 1.738 to account for a difference in normalization between our measurements and those of the HARPS team, and exclude two spectra which appear to show flare events\footnote{BJD = 2453367.7, 2453814.5}.  While the observed slope ($(8 \pm 4) \times 10^{-6}$~day$^{-1}$) is only significant at the $2\sigma$ level, it is entirely consistent with the measured HRS trend.

\section{\bf Radial Velocity}
\label{sec:rv}

GJ 176 has been monitored extensively by HRS, HARPS, and HIRES, resulting in a robust RV data set for the star.  In Figure \ref{gj176_allrv}, we show all 196 available velocities for the star.

\subsection{The HRS and HIRES data}

Because the HRS and HIRES RV series have similar baselines, precisions, and results, we will discuss them together before comparing/contrasting with the HARPS RVs.

As GJ 176 is relatively nearby ($d = 9.3$ pc), its secular acceleration of $\sim 0.4$ m s$^{-1}$ yr$^{-1}$ has become significant over our 10-year observational baseline.  Upon subtracting this acceleration \citep[following][]{zechmeister09} from our RVs\footnote{\citet{forveille09} subtracted secular acceleration from the HARPS velocities, but \citet{butler09} did not subtract it from the HIRES data.  We have therefore removed the secular acceleration from the HIRES velocities.}, we see that a positive linear trend still remains in our velocities.  Plotting RV as a function of \id (Figure \ref{gj176_narv}), we see from the correlation between the two variables that this slope appears to be the RV signature of the stellar activity trend.  Removing the linear fit from the relation eliminates the trend from the RVs, suggesting the acceleration is most likely caused by a long-term magnetic cycle rather than an unseen binary companion.

\begin{figure*}
\begin{center}
\includegraphics[width=1.55\columnwidth]{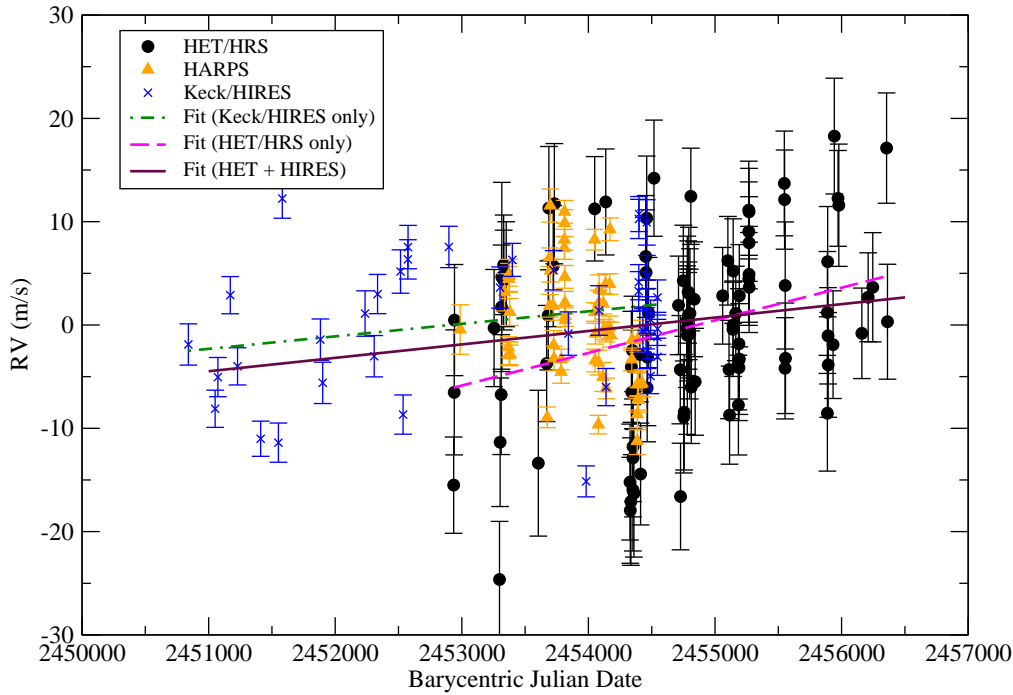}
\caption{All available RVs for GJ 176.  The solid line shows our best linear fit to the combined HRS+HIRES data set, while the dashed pink and dot-dashed green lines show fits to the HRS and HIRES RVs, respectively.  Secular acceleration has been subtracted from the velocities, and we have fit and subtracted a constant offset for each data set.}
\label{gj176_allrv}
\end{center}
\end{figure*}

\begin{figure*}
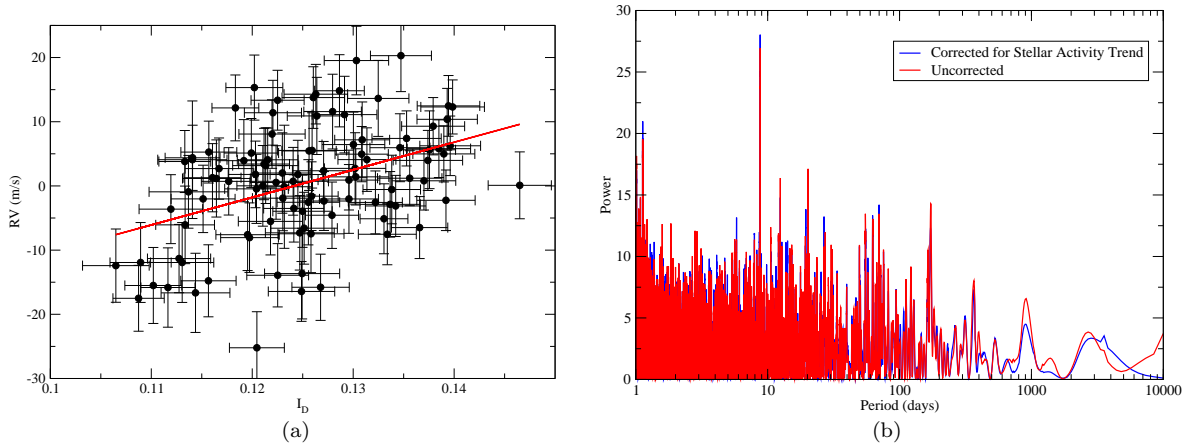

\begin{center}
\subfigure[\label{gj176_narv}]{\includegraphics[width=0.9\columnwidth,trim=0 0 -1.5cm 0]{gj176_narv.eps}}
\subfigure[\label{gj176_pscomp}]{\includegraphics[width=0.9\columnwidth]{gj176_noharps_pscomp.eps}}
\caption{\emph{a}. RV as versus \id for our HET/HRS RVs of GJ 176.  The red line gives our linear least squares fit to the relation.  \emph{b}. Generalized Lomb-Scargle periodogram of the HRS+HIRES RVs of GJ 176 before (red) and after (blue) subtracting Equation \ref{gj176_rvtrend}.  Note: for each of these plots, the secular acceleration has been subtracted from the velocities.}
\label{gj176_actrv}
\end{center}
\end{figure*}

\citet{butler09} indicated the possibility of a linear trend in their Keck/HIRES velocities of GJ 176 which, if confirmed, would add confidence to our own detection.  Examining all the available velocities (Figure \ref{gj176_allrv}), while the data show considerable scatter--as expected from the planetary and rotation signals--a positive linear trend is indeed present over the entire RV series.  Specifically, we find a linear least squares fit of

\begin{equation}
\label{gj176_rvtrend}
\textrm{RV} (\textrm{m s}^{-1}) = -1.1(0.8) + 0.0013(0.0005)t
\end{equation}

Here, $t = \textrm{BJD} - 2\,456\,300$.  This fit allows for a zero-point offset between the two data sets.  We obtain a Pearson correlation coefficient $r = 0.2$ for the RV trend which, for a set of 141 total RVs, gives a probability $P(r) = 0.009$ of a flat slope.  Fitting to the HRS or HIRES sets individually yields fits consistent to within the uncertainties with Equation \ref{gj176_rvtrend}.  We include all three fits in Figure \ref{gj176_allrv}.

Although the high statistical significance of the \id trend and its correlation with our HET/HRS velocities convinces us the overall RV set should contain a slope, because of the large scatter in the RVs, the shallow slope derived from the combined data is only a $\sim2.5\sigma$ detection.  In addition to causing large deviations from the fit, the scatter increases the uncertainties in the zero-point offsets between the RV sets, which can have a large effect on the linear fit.  In an attempt to remedy this problem, we attempted computing Equation \ref{gj176_rvtrend} using the residual RVs around a fit to the planet.  In addition to reducing the velocity scatter, fitting and removing the exoplanet signal offers a more reliable way to determine the zero-point offsets, since the presence of a slope in the RVs could otherwise lead to incorrect results.  Reassuringly, our fit to the slope is identical regardless of whether or not we subtract the planet.  This result, coupled with the fact that removing the slope from the RVs increases the significance of previously-known signals (see below), lends additional evidence to the veracity of Equation \ref{gj176_rvtrend}.

Because we do not have \id values from HIRES, we cannot verify the RV-\id correlation for the HIRES data.  However, given the agreement of the HIRES slope with that seen in the HRS data (which we know is correlated with activity), we strongly suspect the observed trend is due to stellar magnetic activity.  With activity-induced RV contributions from stellar rotation \emph{and} a long-term magnetic cycle, GJ 176 joins $\alpha$ Cen B among stars with multiple stellar RV signals and (candidate) planets.  

While the rotation and exoplanet signals add scatter and inflate the uncertainties on the RV-\id relation (or its manifestation as an RV trend), it is nevertheless tempting to attempt an activity correction for this star to evaluate how the remaining RV signals respond.  In Figure \ref{gj176_pscomp}, we show periodograms of the combined HRS+HIRES RVs before and after subtracting Equation \ref{gj176_rvtrend}.  The power of the peak corresponding to the planet ($P = 8.78$ days) increases from 27 to 28 after subtracting the trend which, assuming the false alarm probability for power $Z$ scales as $e^{-Z}\sqrt{Z}$, translates to a reduction in FAP by a factor of $\sim3$.

\subsection{The HARPS data}

At first glance, our results for both \id and RV in GJ 176 appear to contradict those from the HARPS survey.  However, looking at all the available data for the star with careful considerations for instrumental precision and time baselines shows the findings presented here are not necessarily incompatible with previous studies.

\citet{gds12} list GJ 176 as inactive in \id because it does not pass their variability test, but this result can be attributed mostly to their sampling and analysis methods.  Their data for the star consists of just 7 nights of observations, each binned into a nightly average.  The approach is designed to search for long-term variability and RV correlation.  We therefore do not expect the HARPS data to show rotationally-induced modulation, as such behavior is intentionally ignored.  

Interestingly, although the combined RV set shows the 39-day stellar rotation signal at statistically significant power, when examining each individual data set we find that the HARPS velocities are the only ones which contain \emph{any} periodogram power at the rotation period.  We attribute this feature to the fact that the HARPS observations have a much more frequent cadence, and therefore have a higher probability of resolving a rotation signal before it changes phase.  Additionally, since HARPS has a much higher resolving power ($R \sim$~110\,000) than HRS and HIRES ($R \sim$~50\,000-60\,000), it may be more sensitive to spot-induced modulation of the stellar line profiles, leading to a more robust detection of the rotation period.  The relatively short time baseline prevents phase shifts from degrading the rotation signal, and the relatively high precision causes the signal to persist in the combined RV set since generalized Lomb-Scargle periodograms and $\chi^2$ minimization algorithms give added weight to the HARPS RVs.

It is important to note that the absence of the rotation signal from the HRS/HIRES RVs \emph{cannot} be attributed to RV precision alone.  Performing a 2-signal RV fit to the HARPS velocities alone, we find an RV amplitude of 4.4 \mps for the rotation signal, compared to 4.0 \mps for the planet signal.  As evidenced by the strong detection of the planet in the combined HRS/HIRES data, if the rotation signal remained constant in phase and amplitude, it should have been easily recovered.

We are unable to confidently determine the presence of the long-term RV slope for the HARPS RVs.  In both the \citet{forveille09} and \citet{gds12} RV sets, the raw velocities actually show a \emph{negative} slope, seemingly contradicting the HRS/HIRES slopes.  However, the residuals to a planet-plus-rotation model appear flat.  Evidently, the planet and (especially) the rotation signal dominate the RVs during this period, and have a strong effect on any long-term trends, or lack thereof.  Thus, although the HARPS \id values are consistent with the positive activity trend, we are unable to confirm whether this consistency extends to the RVs due to the mitigating factors of the short time baseline and the periodic signals.

The HARPS data for GJ 176 create a conundrum for this analysis.  The HARPS RVs are the most precise of the three data sets, and the dense time sampling is ideal for resolving the signals of the planet and the stellar rotation.  On the other hand, the relatively short time baseline and the unusually large RV contribution from stellar rotation make it difficult to compute a unified model of the planet, the stellar rotation, and the long-term trend for the combined HRS+HIRES+HARPS RVs.  Even if the rotation-induced RVs were present throughout the observational baseline, phase shifts of the rotation signal would likely prevent a ``global" 2- or 3-signal model.  Since we are primarily interested in the magnetic activity of GJ 176 in this study, we have elected not to pursue a combined RV fit for this reason.

\section{\bf Discussion}

Counting the stellar rotation period, its first harmonic, the long-term trend and the two intermediate-period signals, GJ 176 exhibits at least five distinct stellar activity signals.  Two (rotation and the trend) have already been observed to manifest in RV, and it is reasonable to expect that at the RV amplitudes ($\leq 1$ \mpsp) of potential additional terrestrial planets in the system, the other activity-related periodicities will appear.  Thus, while GJ 176 is certainly an interesting candidate system for discovering and characterizing low-mass exoplanets, its activity makes the RV interpretation especially difficult.

With multiple activity-induced RV signatures, GJ 176 has a reputation as a highly active \mdp.  However, it is important to note that with a mean \lha of -3.84, GJ 176 has an overall stellar activity level that is essentially average for its mass \citep[according to the relation in][]{robertson13a}.  Its rotation period of 39 days is likewise ordinary for an old M star, as opposed to the shorter rotation periods of more active \mdsp.  Rather, what is remarkable about GJ 176 is the fact that its major spots have survived for at least the 6 years covered by our photometric observations, resulting in the coherent photometric signal.  Our results seem to suggest that atypically strong magnetic fields (such as would produce abnormally large mean activity levels) are not required to preserve such spots.  

It is possible that individual spots need not survive for extended periods in order to create the coherent photometric signal we observe.  On the Sun, spots preferentially appear at so-called ``active longitudes," where increased magnetic activity in a localized region causes spots to manifest repeatedly \citep[e.g.][]{berdyugina03,ivanov07}.  Active longitudes rotate in phase with the stellar rotation (modulo differential rotation), and could explain a persistent coherent starspot signal.  However, \citet{ivanov07} finds that prominent solar active longitudes tend to survive for about 20 rotations or less.  If the $\sim$6-year photometric signal of GJ 176 is caused by a single active longitude, it has survived at least 50 rotations, suggesting a significant departure from solar behavior.  More detailed work is required to understand how such long-lasting active regions might be maintained on low-mass stars.

The physical origins of the $\sim$73- and $\sim$112-day activity signals are of some interest, as they are not simple harmonics of the rotation period.  However, they are close to 2 and 3 times the rotation, so it is possible the signals are physically related to rotation.

A study of starspots on the M4 dwarf GJ 1243 \citep{davenport14} offers insight towards a potential explanation.  The \kep lightcurve of GJ 1243 shows sinusoidal variations induced by stellar rotation remaining in phase over more than 4 years, while a secondary ``shoulder" in the lightcurve changes in phase on $\sim100$-day timescales.  These features may be explained by one or more major spots (or perhaps an active region/longitude) persisting on the stellar surface over many rotations, while small, short-lived spots change over the 100-day timescale.

In the case of GJ 176, we propose a similar scenario.  One or more large, highly-spotted active regions must survive over many years, dominating the photometric variability and leading to the observed in-phase rotation signal in the photometry.  In addition to the active regions, many ``minor" spots can appear at variable times and latitudes outside the active regions, also tracing the stellar rotation, but not in phase over long timescales.  The 73- and 112-day (or ~2-3 rotation) periods would then represent typical minor spot lifetimes.

So why does the stellar rotation signal remain in phase in RV and photometry, but not in the absorption-line fluxes?  We suggest this is due to the fact that the photometry traces starspots (and thus rotation) directly, whereas the absorption-line indices trace emission due to the magnetic activity producing the spots.  If the magnetic field across the stellar surface is changing on the timescale of the individual spot lifetimes, then those changes may dominate over localized active longitudes in the spectral activity indices.

Our activity-RV analysis confirms the planetary nature of the 8.8-day periodicity.  The planet's period is not an integer-ratio harmonic of the rotation period, remains in phase, and does not appear in any of our activity tracers, leaving no reasonable suspicion that the signal is produced by activity.  On the other hand, the phase-shifting of the 39-day signal in RV, \iha and \id ensures it is in fact the rotation period, and not produced by magnetic interaction between GJ 176 and a second, more distant planet \citep[a possibility mentioned by][]{forveille09}.

GJ 176 joins a growing number of systems with RV-detected exoplanets and long-period magnetic cycles \citep[e.g.][]{dumusque11,dumusque12,robertson13b}.  In general, activity cycles which create RV signals--presumably via changing magnetic inhibition of convection--appear to be common across a wide range of spectral types \citep{lovis11,gds12,robertson13a}.  These cycles may either mimic exoplanets, or significantly alter the measured orbital properties of real planets, so it is important to properly diagnose and correct them when characterizing RV systems.

The rotation-induced RV signals of GJ 176 again illustrate both the difficulty and importance of careful activity analysis and correction when attempting to identify low-mass planets in the habitable zones of \mdsp.  Recent analyses of the quiet M stars GJ 581 \citep{robertson14a} and GJ 667C \citep{robertson14b} reinforced the prediction of \citet{boisse11} that activity-induced RV signals appear at the stellar rotation period and its harmonics.  For slowly rotating ($P_{rot} \sim 100$d) M stars, these rotation harmonics are coincident with the periods of planets in the habitable zone.  For GJ 176, where the relatively faster rotation period shifts most of its harmonics inward of the HZ, we see additional activity signals at intermediate periods.  If these signals also appear in RV at the 0.5-1 \mps amplitudes expected for both activity signals and super-Earths in the HZ, it will again create confusion when searching for planets of astrobiological interest.  Indeed, the 39-day rotation period and the 112-day activity signal roughly enclose the optimistic habitable zone of GJ 176 according to \citet{kopparapu13}.  It is therefore clearly possible for the period space of an M star's habitable zone to be completely contaminated with stellar activity signals, even when the star rotates quickly enough that its harmonics are at shorter periods.

\section{\bf Summary}

We have presented a long-term study of magnetic activity of the M dwarf GJ 176 based on spectral activity indicators and optical photometry.  In addition to the stellar rotation period and its first harmonic, we identify two intermediate-period activity signals at periods close to 2 and 3 times the stellar rotation, and a linear trend indicative of a long-period magnetic cycle.  The magnetic cycle and the stellar rotation appear in RV, although the rotation signal is only detectable in the HARPS RVs, which cover only a small fraction of the time baseline provided by our data and the HIRES RVs.  Our results lend additional confirmation that the sodium resonance lines are a powerful tool for identifying and characterizing activity-induced RV shifts in M stars, as they are sensitive to the magnetic cycle and the 70-day periodicity, and correlate with RV, whereas \ha does not.

\begin{acknowledgements}
We thank the anonymous referee for his or her suggestions for improving this paper.  P. R. gratefully acknowledges support from the National Science Foundation via MRI grant number 1126413, the Penn State Center for Exoplanets and Habitable Worlds postdoctoral fellowship, and the University of Texas graduate continuing fellowship.  This work was also supported by the National Aeronautics and Space Administration through the Origins of Solar Systems Program grant NNX09AB30G and grant AST \#1313075 from the NSF.  Previous support for the McDonald Observatory exoplanet survey was given through NASA grants NNX07AL70G and NNX10AL60G.  The Hobby-Eberly Telescope (HET) is a joint project of the University of Texas at Austin, the Pennsylvania State University, Stanford University, Ludwig-Maximilians-Universit\"{a}t M\"{u}nchen, and Georg-August-Universit\"{a}t G\"{o}ttingen.  The HET is named in honor of its principal benefactors, William P. Hobby and Robert E. Eberly.  We would like to thank the McDonald Observatory TAC for generous allocation of observing time.  We are grateful to the HET Resident Astronomers and Telescope Operators for their valuable assistance in gathering our HET/HRS data.
\end{acknowledgements}


{}

\end{document}